\documentclass[twocolumns,jgrga]{agu2001}
\usepackage{graphicx}
 \providecommand\bcdot{{\mathbf
\cdot}}

\authorrunninghead{MAGNE ET AL.}

\titlerunninghead{Waves over submarine canyons}

\authoraddr{Fabrice Ardhuin,
Centre Militaire d'Oc{\'e}anographie, Service Hydrographique et
Oc\'{e}anographique de la Marine, 29609 Brest, France. (ardhuin@shom.fr)}

\begin{document}
\setkeys{Gin}{draft=false}

\title{Evolution of surface gravity waves over a submarine canyon}

\authors{R. Magne  \altaffilmark{1,5},
  K. A. Belibassakis \altaffilmark{2},
  T. H. C. Herbers \altaffilmark{3},
Fabrice Ardhuin \altaffilmark{1}, \\ W. C. O'Reilly \altaffilmark{4}, and V.
Rey \altaffilmark{5}}

\altaffiltext{1}
{Centre Militaire d'Oc{\'e}anographie, Service Hydrographique et
Oc\'{e}anographique de la Marine, 29609 Brest, France.}

\altaffiltext{2}
{Department of Naval Architecture and Marine Engineering, National Technical
University of Athens, PO Box 64033 Zografos, 15710 Athens, Greece.}

\altaffiltext{3}
{Department of Oceanography, Naval Postgraduate School, Monterey, CA
93943, USA.}

\altaffiltext{4}
{Integrative Oceanography Division, Scripps Institution of Oceanography, La
Jolla, CA 92093, USA.}

\altaffiltext{5}
{Laboratoire de Sondages Electromagn{\'e}tique de l'Environnement Terrestre,
Universit{\'e} de Toulon et du Var, La Garde, France.}

\begin{abstract}
The effects of a submarine canyon on the propagation of ocean surface waves are
examined with a three-dimensional coupled-mode model for wave propagation over
steep topography. Whereas the classical geometrical optics approximation
predicts an abrupt transition from complete transmission at small incidence
angles to no transmission at large angles, the full model predicts a more
gradual transition with partial reflection/transmission that is sensitive to
the canyon geometry and controlled by evanescent modes for small incidence
angles and relatively short waves. Model results for large incidence angles are
compared with data from directional wave buoys deployed around the rim and over
Scripps Canyon, near San Diego, California, during the Nearshore Canyon
Experiment (NCEX). Wave heights are observed to decay across the canyon by
about a factor 5 over a distance shorter than a wavelength. Yet, a spectral
refraction model predicts an even larger reduction by about a factor 10,
because low frequency components cannot cross the canyon in the geometrical
optics approximation. The coupled-mode model yields accurate results over and
behind the canyon. These results show that although most of the wave energy is
refractively trapped on the offshore rim of the canyon, a small fraction of the
wave energy `tunnels' across the canyon. Simplifications of the model that
reduce it to the standard and modified mild slope equations also yield good
results, indicating that evanescent modes and high order bottom slope effects
are of minor importance for the energy transformation of waves propagating
across depth contours at large oblique angles.
\end{abstract}

\begin{article}

\section{Introduction}
Waves are strongly influenced by the bathymetry when they reach shallow
water
areas. \textit{Munk and Traylor} [1947]\nocite{Munk&Traylor1947} conducted a
first quantitative study of the effects of bottom topography on wave energy
transformation over Scripps and La Jolla Canyons, near San Diego,
California.
Wave refraction diagrams were constructed using a manual method,
and compared to visual observations. Fairly good agreement was
found between predicted and observed wave heights.
Other effects such as diffraction were found to be important elsewhere, for
sharp bathymetric features (e.g. harbour structures or coral reefs), prompting
\textit{Berkhoff} [1972]\nocite{Berkhoff1972} to introduce an equation that
represents both refraction and diffraction. Berkhoff's equation is based on a
vertical integration of Laplace's equation and is valid in the limit of small
bottom slopes. It is widely known as the mild slope equation (MSE). A parabolic
approximation of this equation was proposed by \textit{Radder}
[1979]\nocite{Radder1979}, and further refined by \textit{Kirby}
[1986]\nocite{Kirby1986c} and \cite{Dalrymple&Kirby1988}.

\textit{O'Reilly and Guza} [1991,
1993]\nocite{OReilly&Guza1991,OReilly&Guza1993} compared \textit{Kirby}'s
[1986] refraction-diffraction model to a spectral geometrical optics refraction
model based on the theory of \textit{Longuet-Higgins}
[1957\nocite{Longuet-Higgins1957}]. The two models generally agreed in
simulations of realistic swell propagation in the Southern California Bight.
However, both models assume a gently sloping bottom, and their limitations in
regions with steep topography are not well understood. \textit{Booij}
[1983]\nocite{Booij1983}, showed that the MSE is valid for bottom slopes as
large as $1/3$ for normal wave incidence. To extend its application to steeper
slopes, \textit{Massel} [1993 ; see also \textit{Chamberlain and Porter},
1995]\nocite{Massel1993,Chamberlain&Porter1995} modified the MSE by including
terms of second order in the bottom slope, that were neglected
by\nocite{Berkhoff1972} \textit{Berkhoff} [1972]. This modified mild slope
equation (MMSE) includes terms proportional to the bottom curvature and the
square of the bottom slope. \textit{Chandrasekera and Cheung}
[1997]\nocite{Chandrasekera&Cheung1997} observed that the curvature terms
significantly change the wave height behind a shoal, whereas the slope-squared
terms have a weaker influence. \textit{Lee and Yoon}
[2004]\nocite{Lee&Yoon2004} noted that the higher order bottom slope terms
change the wavelength, which in turn affects the refraction. In spite of these
improvements, an important restriction of these equations is that the vertical
structure of the wave field is described by the Airy solution of waves over a
horizontal bottom. Hence the MMSE cannot describe the wave field accurately
over steep bottom topography. Thus, \textit{Massel} [1993] introduced an
additional infinite series of local modes ('evanescent modes' or 'decaying
waves'), that allows a local adaptation of the wave field [see also
\textit{Porter and Staziker}, 1995]\nocite{Porter&Staziker1995}, and converges
to the exact solution of Laplace's equation, except at the bottom interface.
Indeed, the vertical velocity at the bottom is still zero, and is discontinuous
in the limit of an infinite number of modes. Recently, \textit{Athanassoulis
and Belibassakis} [1999]\nocite{Athanassoulis&Belibassakis1999} added a
'sloping bottom mode' to the local mode series expansion, which properly
satisfies the Neuman bottom boundary condition. This approach was further
explored by \textit{Chandrasekera and Cheung},
[2001]\nocite{Chandrasekera&Cheung2001} and \textit{Kim and Bai},
[2004]\nocite{Kim&Bai2004}. Although the sloping-bottom mode yields only small
corrections for the wave height, it significantly improves the accuracy of the
velocity field close to the bottom. Moreover, this mode enables a faster
convergence of the series of evanescent modes, by making the convergence
mathematically uniform.

As these steep topography models are becoming available, one may wonder if this
level of sophistication is necessary to accurately describe the transformation
of ocean waves over natural continental shelf topography. It is expected that
if such models are to be useful anywhere, it should be around steep submarine
canyons. Surprisingly, a geometrical optics refraction model that assumes weak
amplitude gradients on the scale of the wavelength, usually corresponding to
gentle bottom slopes, was found to yield accurate predictions of swell
transformation over Scripps canyon [\textit{Peak}, 2004\nocite{Peak2004}]. The
practical limitations of mild slope approximations for natural seafloor
topography are clearly not well established.

The goal of the present paper is to understand the propagation of waves over a
submarine canyon, including the practical imitations of geometrical optics
theory for the associated large bottom slopes. Numerical models will be used to
sort out the relative importance of refraction, and diffraction effects.
Observations of ocean swell transformation over Scripps and La Jolla Canyons,
collected during the Nearshore Canyon Experiment (NCEX), are compared with
predictions of the three-dimensional (3D) coupled-mode model. This model is
called NTUA5 because its present implementation will be limited to a total of 5
modes [\textit{Belibassakis et al.}, 2001\nocite{Belibassakis&al.2001}]. This
is the first verification of a NTUA-type model with field observations, as
previous model validations were done with laboratory data. This application of
NTUA5 to submarine canyons is not straightforward since the model is based on
the extension of the two-dimensional (2D) model of
\cite{Athanassoulis&Belibassakis1999}, and requires special care in the
position of the offshore boundary and the numerical damping of scattered waves
along the boundary. Further details on these and software developments, and a
comparison with results of the SWAN model [\textit{Booij et al.},
1999\nocite{Booij&al.1999}] for the same NCEX case are given by
\cite{Gerosthathis&al.2005}.

Here, model results are compared with two earlier models which assume a gently
sloping bottom. These are the parabolic refraction/diffraction model REF/DIF1
(V2.5) [\textit{Kirby}, 1986], applied in a spectral sense, and a spectral
refraction model based on backward ray tracing [\textit{Dobson},
1967\nocite{Dobson1967} ; \textit{O'Reilly and Guza}, 1993]. A brief
description of the coupled-mode model and the problems posed by its
implementation in the NCEX area is given in section 2. Although our objective
is the understanding of complex 3D bottom topography effects in the NCEX
observations, this requires some prior analysis, performed in section 3, of
reflection and refraction patterns over idealized 2D canyons. Results are
presented for realistic transverse canyon profiles, including a comparison with
the 2D analysis of infragravity wave observations reported by
\cite{Thomson&al.2005}. Comparisons of 3D models with field data are presented
in section 4 for representative swell events observed during NCEX. Conclusions
follow in section 5.

\section{Numerical Models}
The fully elliptic 3D model developed by \textit{Belibassakis et al.} [2001] is
based on the 2D model of \textit{Athanassoulis and Belibassakis} [1999]. These
authors formulate the problem as a transmission problem in a finite subdomain
of variable depth $h_2(x)$ (uniform in the lateral y-direction), closed by the
appropriate matching conditions at the offshore and inshore boundaries. The
offshore and inshore areas are considered as incidence and transmission regions
respectively, with uniform but different depths ($h_1$, $h_3$), where complex
wave potential amplitudes $\varphi_1$ and $\varphi_3$ are represented by
complete normal-mode series containing the propagating and evanescent modes.


The wave potential $\varphi_2$ associated with $h_2$ (region $2$),
is given by the following local mode series expansion:
\begin{eqnarray}
\varphi_2(x,z)  &= &
\varphi_{-1}(x)Z_{-1}(z;x)+\varphi_{0}(x)Z_{0}(z;x)\nonumber\\
        & & +   \sum_{n=1}^{\infty}\varphi_n(x)Z_n(z;x),
        \label{phi}
\end{eqnarray}


where $\varphi_{0}(x)Z_{0}(z;x)$ is the propagating mode and
$\varphi_n(x)Z_n(z;x)$ are the evanescent modes. The additional term
$\varphi_{-1}(x)Z_{-1}(z;x)$ is the sloping-bottom mode, which permits the
consistent satisfaction of the bottom boundary condition on a sloping
bottom.
The modes allow for the local adaptation of the wave potential. The
functions
$Z_n(z;x)$ which represent the vertical structure of the $n^{\mathrm{th}}$
mode
are given by,
\begin{equation}\label{Z0}
Z_0(z,x)=\frac{\cosh[k_0(x)(z+h(x))]}{\cosh(k_0(x)h(x))},
\end{equation}
\begin{equation}\label{Zn}
Z_n(z,x)=\frac{\cos[k_n(x)(z+h(x))]}{\cos(k_n(x)h(x))},~~~n=1,2,...,
\end{equation}
\begin{equation}\label{Z-1}
Z_{-1}(z,x)=h(x)\left[\left(\frac{z}{h(x)}\right)^3 +
\left(\frac{z}{h(x)}\right)^2 \right],
\end{equation}
where $k_0$ and $k_n$ are the wavenumbers obtained from the
dispersion relation (for propagating and evanescent modes),
evaluated for the local depth $h=h(x)$:
\begin{equation}\label{disprelpropa}
\omega^2=gk_0\tanh{k_0h}=-gk_n\tan{k_nh},
\end{equation}
with $\omega$ the angular frequency

As discussed in \textit{Athanassoulis and Belibassakis} [1999], alternative
formulations of $Z_{-1}$ exist, and the extra sloping-bottom mode controls only
the rate of convergence of the expansion (\ref{phi}) to a solution that is
indeed unique. The modal amplitudes $\varphi_n$ are obtained by a variational
principle, equivalent to the combination of Laplace's equation, the bottom and
surface boundary conditions, and the matching conditions at the side
boundaries, leading to the coupled-mode system,
\begin{eqnarray}
\label{coupled-mode system}
\sum_{n=-1}^{\infty}a_{mn}(x)\varphi''_n(x)&+&b_{mn}(x)\varphi'_n(x)+
c_{mn}(x)\varphi_n(x)=0,\nonumber
\\ & & \mathrm{for}\quad(m=-1,0,1,...)
\end{eqnarray}
where $a_{mn}$, $b_{mn}$ and $c_{mn}$ are defined in terms of the $Z_n$
functions, and the appropriate end-conditions for the mode amplitudes
$\varphi_n$ ; for further details, see \textit{see Athanassoulis and
Belibassakis} [1999]. The sloping-bottom mode ensures absolute and uniform
convergence of the modal series. The rate of decay for the modal function
amplitude is proportional to ($n^{-4}$). Here, the number of evanescent modes
is truncated at $n=3$, which ensures satisfactory convergence, even for bottom
slopes exceeding 1.

This 2D solution is further extended to realistic 3D bottom topographies by
\textit{Belibassakis et al.} [2001]. In 3D, the depth $h_2$ is decomposed into
a background parallel-contour surface $h_i(x)$ and a scattering topography
$h_d(x,y)$. The 3D solution is then obtained as the linear superposition of
appropriate harmonic functions corresponding to these two topographies. There
is no limitation on the shape and amplitude of the bottom represented by
$h_d(x,y)$ except that $h_d > 0$, which can always be enforced by a proper
choice of $h_i$, for further details see \textit{Belibassakis et al.} [1999].
The wave potential solution over the 2D topography ($h_i$) is governed by the
equations described previously. The wave potential associated with the
scatterers ($h_d$) is obtained as the solution of a 3D scattering problem. The
decomposition of the topography in $h_d$ and $h_i$ is not uniquely defined by
the constraints that $h_i$ is invariant along $y$ and $h_d > 0$, and there is
thus no simple physical interpretation of the scattered field which corresponds
to both reflection and refraction effects. The main benefit of this
decomposition is that the scattered wave field propagates out of the model
domain along the entire boundary, which greatly simplifies the specification of
the horizontal boundary conditions.

In practice we chose
\begin{equation}
h_i(x)= \min \left\{ h(x,y) \quad \mathrm{for} \quad y \in \left[
y_{\mathrm{min}} , y_{\mathrm{max}} \right] \right\}.
\end{equation}
Further, the bathymetry $h_i+h_d$ is modified by including a transition region
for $y< y_{\mathrm{min}}$ and $y > y_{\mathrm{max}}$ in which $h_d$ goes to
zero at the model boundary, so that no scattering sources are on the boundary
and waves actually propagate out of the domain. This modification of the
bathymetry does not change the propagation of the incoming waves, provided that
the offshore boundary is in uniform water depth, as in the cases described by
\cite{Belibassakis&al.2001}, or in deep enough water so that a uniform water
depth can be prescribed without having an effect on the waves. Solutions are
obtained by solving a coupled-mode system, similar to Eq.(\ref{coupled-mode
system}), but extended to two horizontal dimensions $(x,y)$, and coupled with
the boundary conditions ensuring outgoing radiation. The spatial grid for the
scattered field is extended with a damping layer all around the boundary
[\textit{Belibassakis et al.}, 2001].

Both $2$D and $3$D implementations of this NTUA5 model are used here to
investigate wave propagation over a submarine canyon. If we neglect the
sloping-bottom mode and the evanescent modes, and retain in the local-mode
series only the propagating mode $\varphi_0(x,y)$, this model (NTUA5) exactly
reduces to MMSE [e.g. \textit{Chandrasekara and Cheung}, 1997],
\begin{eqnarray}\label{MMS}
\nabla^2\varphi_0(x,y)&+&\frac{\nabla(CC_g)}{CC_g}\bcdot\nabla\varphi_0(x,y)\nonumber
\\ &+&\left[k_0^2+f_1 \nabla^2 h+f_2 (\nabla h)^2\right] \varphi_0(x,y)
=0,
\end{eqnarray}
where $f_1=f_1(x,y)$ and $f_2=f_2(x,y)$ are respectively functions
dependent on the bottom curvature and slope-squared terms. From
Eq.(\ref{MMS}), the MSE is obtained by further neglecting the curvature and
slope-squared terms.

In the following sections, these two formulations (MSE and MMSE) will be
compared to the full 5-mode model to examine the importance of steep bottom
slope effects, which are fully accounted for in this model. The MSE and MMSE
solutions are obtained by exactly the same scattering method described above
with the same computer code in which the high order bottom slope terms
and/or
evanescent modes are turned off. For 3D calculations, our use of a regular
grid
sets important constraints on the model implementation due to the
requirements
to have the offshore boundary in deep water and sufficient resolution to
resolve the wavelength of waves in the shallowest parts of the model domain.
These constraints put practical limits on the domain size for a given wave
period and range of water depths. Here a minimum of 7 points per wavelength
in
10~m depth was enforced, in a domain that extends 4--6~km offshore. Such a
large domain with a high resolution leads to memory intensive inversion of
large sparse matrices. However, the NTUA, MSE and MMSE models are linear,
and
thus the propagation of the different offshore wave components can be
performed
separately, sequentially or in parallel.

Before considering the full complexity of the 3D Scripps-La Jolla Canyon
system, we first examine the behavior of these models in the case of
monochromatic waves propagating over $2$D idealized canyon profiles (transverse
sections of the actual canyons). We consider both the relatively wide La Jolla
Canyon where infragravity wave reflection was reported recently
[\textit{Thomson et al.} 2005], and the narrow Scripps Canyon, that was the
focus of the NCEX swell propagation study.

\section{Idealized 2D canyon profiles}

\subsection{Transverse section of La Jolla Canyon}
\begin{figure}[htbp]
\centerline{\includegraphics[width=20pc]{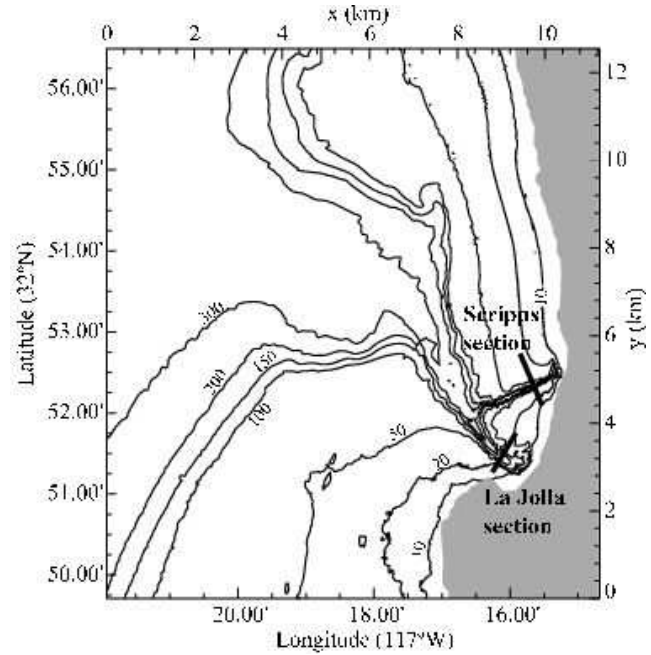}}
\caption{Bathymetry around La Jolla and Scripps canyons, and definition of
transverse sections for idealized calculations.} \label{2sections}
\end{figure}

\begin{figure}[htbp]
\centerline{\includegraphics[width=20pc]{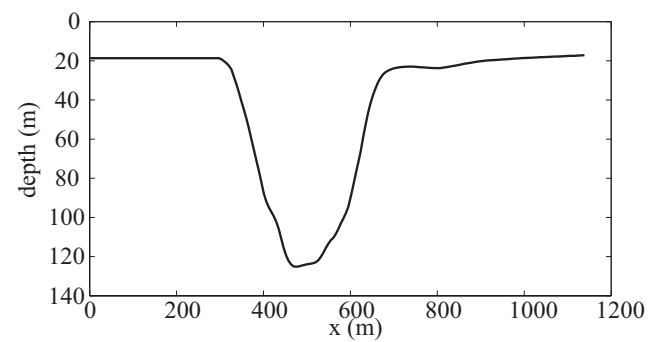}}
\caption{Water depth across the La Jolla canyon section.} \label{depthsouth}
\end{figure}

We investigate monochromatic waves propagating at normal incidence over a
transverse section of the La Jolla Canyon (Figures
\ref{2sections},\ref{depthsouth}), which is relatively deep (120~m) and wide
(350~m). Oblique incidence will not be considered for this canyon because the
results are similar to those obtained for Scripps Canyon (discussed below).
Reflection coefficients $R$ for the wave amplitude are computed using the MSE,
the MMSE, and the full coupled-mode model NTUA5. $R$ is easily obtained using
the natural decomposition provided by the scattering method, and is defined as
the ratio between the scattered wave potential amplitude, up-wave of the
topography, and the amplitude of the imposed propagating wave. In addition, a
stepwise bottom approximation model developed by \textit{Rey}
[1992]\nocite{Rey1992}, based on the matching of integral quantities at the
boundaries of adjacent steps, is used to evaluate $R$ [see \textit{Takano},
1960\nocite{Takano1960}; \textit{Miles}, 1967\nocite{Miles1967}; \textit{Kirby
and Dalrymple}, 1983]\nocite{Kirby&Dalrymple1983}. This model is known to
converge to the exact value of $R$, and will be used as a benchmark for this
study.
\begin{figure}[htbp]
\centerline{\includegraphics[width=20pc]{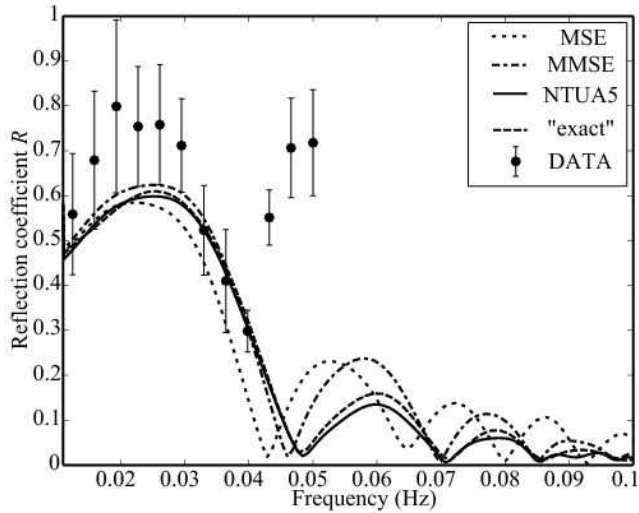}}
\caption{Amplitude reflection coefficient $R$ for waves propagating at normal
incidence over the La Jolla canyon section (figure \ref{depthsouth}) using
several numerical models, and observed infragravity reflections for near-normal
incidence angles [\textit{Thomson et al.}, 2005]}. \label{KRRprofilesouth}
\end{figure}
\begin{figure}[htbp]
\centerline{\noindent\includegraphics[width=20pc]{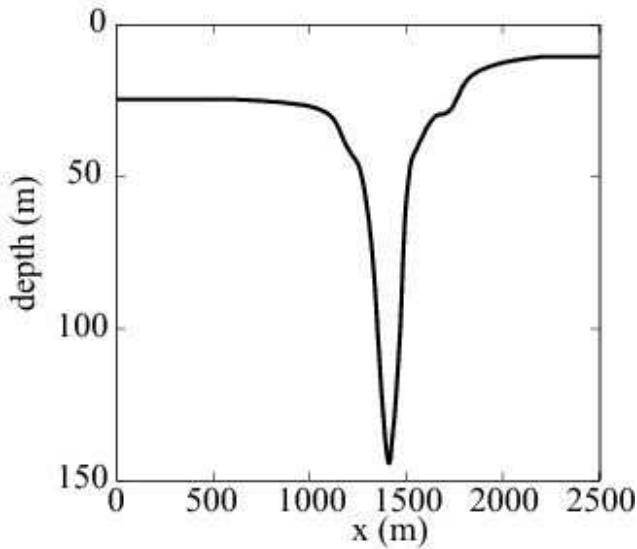}}
\caption{Water depth across the Scripps canyon section.} \label{depth3ral}
\end{figure}
The canyon profile is resolved with 70 steps which was found to be sufficient
to obtain a converging result. The predicted values of $R$ as a function of
wave frequency $f$ (Figure \ref{KRRprofilesouth}), are characterized by maxima
and minima, which are similar to the rectangular step response shown in
\textit{Mei and Black} [1969]\nocite{Mei&Black1969}, \textit{Kirby and
Dalrymple} [1983a], and \textit{Rey et al.} [1992]\nocite{Rey&al.1992}. The
spacing between the minima or maxima is defined by the width of the step or
trench, which imposes resonance conditions, leading to constructive or
destructive interferences. Both the MSE and MMSE models are found to generally
overestimate the reflection at high frequencies, whereas the NTUA5 model is in
good agreement with the benchmark solution. The sloping-bottom mode included in
NTUA5 has a negligible impact on the wave reflection in this and other cases
discussed below. The only other difference between the NTUA5 and the MMSE
models is the addition of the evanescent modes which, through their effect on
the near wave field solution modify significantly the far field, including the
overall reflection and transmission over the canyon.

\cite{Thomson&al.2005} investigated the transmission of infra-gravity waves
with frequencies in the range $0.006$--$0.05$~Hz across this same canyon. Based
on pressure and velocity time series at two points located approximately at the
ends of the La Jolla section these authors estimated energy reflection
coefficients as a function of frequency. In a case of near-normal incidence
they observed a minimum of wave reflection at about 0.04~Hz, generally
consistent with the present results (figure \ref{KRRprofilesouth}).
\cite{Thomson&al.2005} further found a good fit of their observations to the
theoretical reflection across a rectangular trench as given by
\cite{Kirby&Dalrymple1983} in the limit of long waves, and neglecting
evanescent modes. This approximation is appropriate for the long infragravity
band for which the effects of evanescent modes are relatively weak. The
observations of \cite{Thomson&al.2005} also agree well with the various models
applied here to the actual canyon profile (figure \ref{KRRprofilesouth}). At
higher swell frequencies ($f > 0.05$~Hz), the MSE, MMSE and NTUA model results
diverge for normal incidence (figure \ref{KRRprofilesouth}). However, contrary
to the beach-generated infragravity waves, swell arrives from the open ocean
and thus always reaches this canyon with a large oblique angle, for which the
differences between these models are small (not shown).

\subsection{Transverse section of Scripps Canyon}
\subsubsection{Normal incidence}
The north branch of the canyon system, Scripps Canyon, provides a very
different effect due to a larger depth ($145$ m) and a smaller width ($250$ m).
Scripps Canyon is also markedly asymmetric with different depths on either
side. A representative section of this canyon is chosen here (Figure
\ref{depth3ral}). The bottom bottom slope locally exceeds 3, i.e. the bottom
makes an angle up to $70^{\circ}$ with the vertical. Reflection coefficient
predictions for waves propagating at normal incidence over the canyon section
are shown in Figure \ref{KRRprofile3ral}. $R$ decreases with increasing
frequency without the pronounced side lobe pattern predicted for the La Jolla
Canyon section. Again, the NTUA5 results are in excellent agreement with the
exact solution. The MSE dramatically underestimates $R$ at low frequencies, and
overestimates $R$ at high frequencies. However, the MMSE is in fairly good
agreement with the benchmark solution in this case, suggesting that the higher
order bottom slope terms are important for the steep Scripps Canyon profile
reflection, while the evanescent modes play only a minor role.

\subsubsection{Oblique incidence}\label{Oblique incidence}
The swell observed near Scripps Canyon generally arrives at a large oblique
angle at the offshore canyon rim. To examine the influence of the incidence
angle $\theta_i$, a representative swell frequency $f=0.067$~Hz was selected,
and the reflection coefficient was evaluated as a function of $\theta_i$. The
amplitude reflection coefficient $R$ is very weak when $\theta_i$ is small, and
as $\theta_i$ increases, $R$ jumps to near-total reflection within a narrow
band of direction around $35^\circ$ (Figure 6).
\begin{figure}[htbp]
\centerline{\noindent\includegraphics[width=19pc]{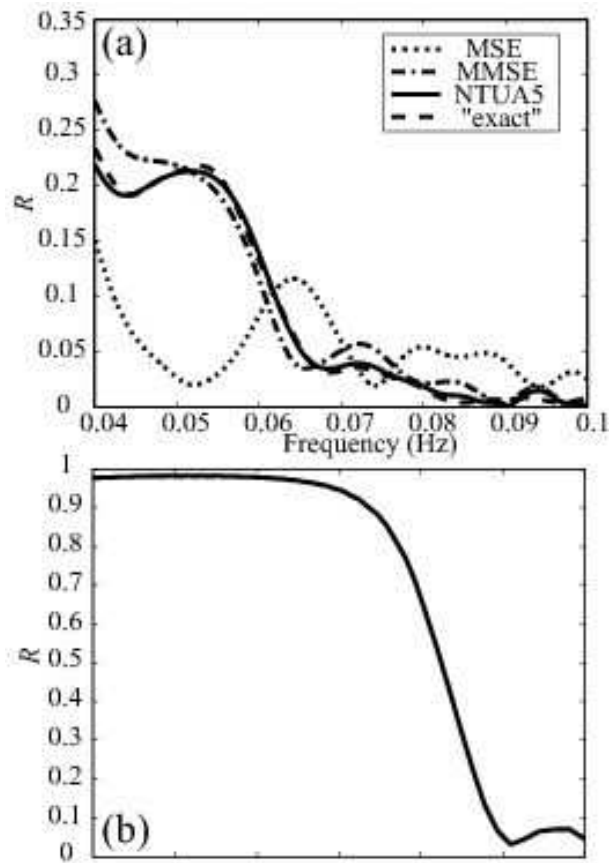}}
\caption{Reflection coefficient for the Scripps Canyon section as a function of
frequency predicted by various models. (a) normal incidence
$\theta_i=0^{\circ}$, (b) $\theta_i=45^{\circ}$. All models collapse on the
same curve in (b).} \label{KRRprofile3ral}
\end{figure}
\begin{figure}[htbp]
\centerline{\noindent\includegraphics[width=20pc]{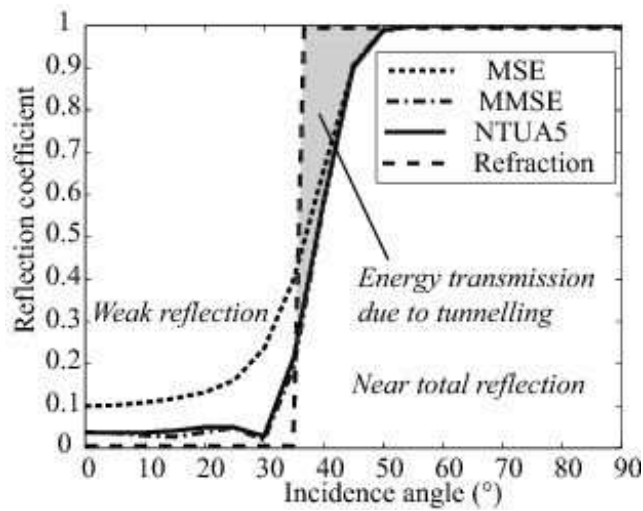}}
\caption{Reflection coefficient for waves of period $T=16$~s propagating over
the Scripps Canyon section as a function of the wave incidence angle $\theta_i$
($0$ corresponds to waves travelling perpendicular to the canyon axis).}
\label{KRR_vs_theta_f_0067}
\end{figure}
Indeed, for a wave train
propagating through a medium with phase speed gradient in one dimension only,
geometrical optics predicts that beyond a threshold (Brewster) angle
$\theta_B$, all the wave energy is trapped, and no energy goes through the
canyon. This sharp transition does not depend on the magnitude of the gradient
which may even be infinite. For a shelf depth $H_1$ and maximum canyon depth
$H_{\mathrm max}$, this threshold angle is given by
\begin{equation}
\theta_B = \arcsin \left(\frac{C_{1}}{C_{\mathrm{max}}}\right)
\end{equation}
where $C_1$ and $C_{\mathrm{max}}$ are the phase speeds for a given frequency
corresponding to the depths $H_1$ and $H_{\mathrm{max}}$. Thus $\theta_B$
increases with increasing frequency as the phase speed difference diminishes at
high frequencies. For Scripps Canyon, $H_1=24$~m, and $H_{\mathrm{max}}=145$~m.
At $f=0.067$~Hz this gives $\theta_B=38^\circ$. As a result, for
$\theta_i<\theta_B$, no reflection is predicted by refraction theory (dashed
line), and all the wave energy is transmitted through the canyon. This
threshold value separates distinct reflection and refraction (trapping)
phenomena, respectively occurring for $\theta_i<\theta_B$ and
$\theta_i>\theta_B$.

The elliptic models that account for diffraction predict a smoother transition.
For $\theta_i<\theta_B$, weak reflection is predicted. For $\theta_i>\theta_B$,
a fraction of the energy is still transmitted through the canyon. This
transmission of wave energy across a deep region where $\sin \theta_i /C_1$
exceeds  $1/C_{\mathrm{max}} $, violates the geometrical optics approximation.
This transmission is similar to the tunnelling of quantum particles through a
barrier of potential in the case where the barrier thickness is of the order of
the wavelength or less [\textit{Thomson et al.}, 2005]. The wave field near the
turning point of wave rays in the canyon decays exponentially in space on the
scale of the wavelength [e.g. \textit{Chao and Pierson},
1972\nocite{Chao&Pierson1972}], and that decaying wave excites a propagating
wave on the other side of the canyon. This coupling of both canyon sides
generally decreases as the canyon width or the incidence angle increase
[\textit{Kirby and Dalrymple}, 1983; \textit{Thomson at al.},
2005]\nocite{Thomson&al.2005}. The significant differences between MSE and MMSE
at small angles $\theta_i<\theta_B$ are less pronounced for
$\theta_i>\theta_B$.

\begin{figure}[htbp]
\centerline{\noindent\includegraphics[width=18pc]{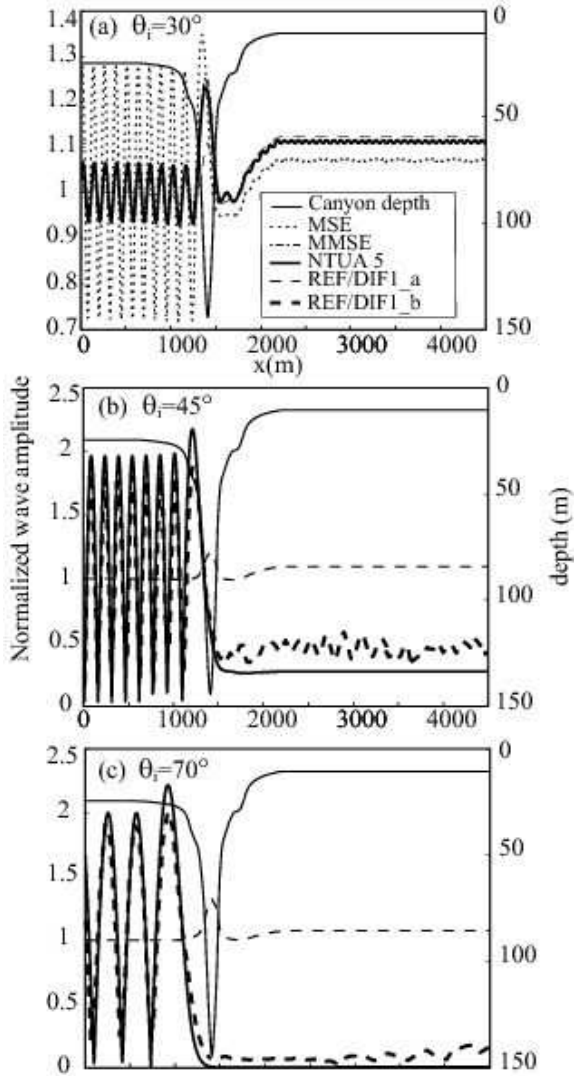}}
\caption{Wave amplitude over the Scripps Canyon section, for $T=16$~s and
different incident angles (a) $\theta_i=30^{\circ}$, (b) $\theta_i=45^{\circ}$,
and (c) $\theta_i=70^{\circ}$. The canyon depth profile is indicated with a
thin dashed line. The MMSE result is in distinguishable from that of NTUA5 in
all panels, and all models except for REF/DIF$1$ give the same results in (b)
and (c).} \label{phiprofile3ralT}
\end{figure}

These two regimes are illustrated by the evolution of the wave potential
amplitude over the Scripps canyon section. In  figure \ref{phiprofile3ralT},
results of various elliptic models (MSE, MMSE and NTUA5) are compared with a
parabolic approximation of the MSE (the REF/DIF1 model of \textit{Dalrymple and
Kirby} [1988]). It should be noted that the model grid orientation is chosen
with the main axis along the incident wave propagation direction, in order to
minimize large angle errors in the parabolic approximation. In that
configuration, the parabolic approximation (REF/DIF1\_a) does not predict any
reflection, but gives an indication of the expected shoaling of the incident
waves across the canyon. For $\theta_i=30^\circ<\theta_B$, weak reflection
(about 10\%) is predicted by the MMSE and NTUA5 (figure
\ref{phiprofile3ralT}.a). MSE considerably overestimates the reflection, and
thus underestimates the transmitted energy down-wave of the canyon section. A
partial standing wave pattern is predicted up-wave of the canyon as a result of
the interference of incident and reflected waves. The largest amplitudes, about
$20\%$ larger than the incident wave amplitude, occur in the first antinode
near the canyon wall.

For a larger wave incidence angle (e.g. 45$^\circ > \theta_B$), an almost
complete standing wave pattern is predicted by the elliptic models up-wave of
the canyon, with an exponential tail that extends across the canyon to a weak
transmitted component (see also Figure \ref{KRRprofile3ral}.b for the
reflection coefficient pattern). Finally, transmission is extremely weak for
$\theta_i=70^\circ$ (figure \ref{phiprofile3ralT}.c). A good estimate of the
reflection coefficient can also be obtained with the parabolic model
REF/DIF1\_b by choosing the x-axis to be aligned with the canyon trench (figure
7b,c thick dashed lines).

\section{West Swell Over Scripps Canyon}
The models used in the previous section (MSE, MMSE, NTUA5,
REF/DIF$1$, refraction) are now applied to the real $3$D bottom
topography of the Scripps-La Jolla Canyon system, and compared
with field data from directional wave buoys deployed around the
rim and over Scripps Canyon during NCEX.

\subsection{Models Set-up}
\begin{figure}[htbp]
\centerline{\noindent\includegraphics[width=19pc]{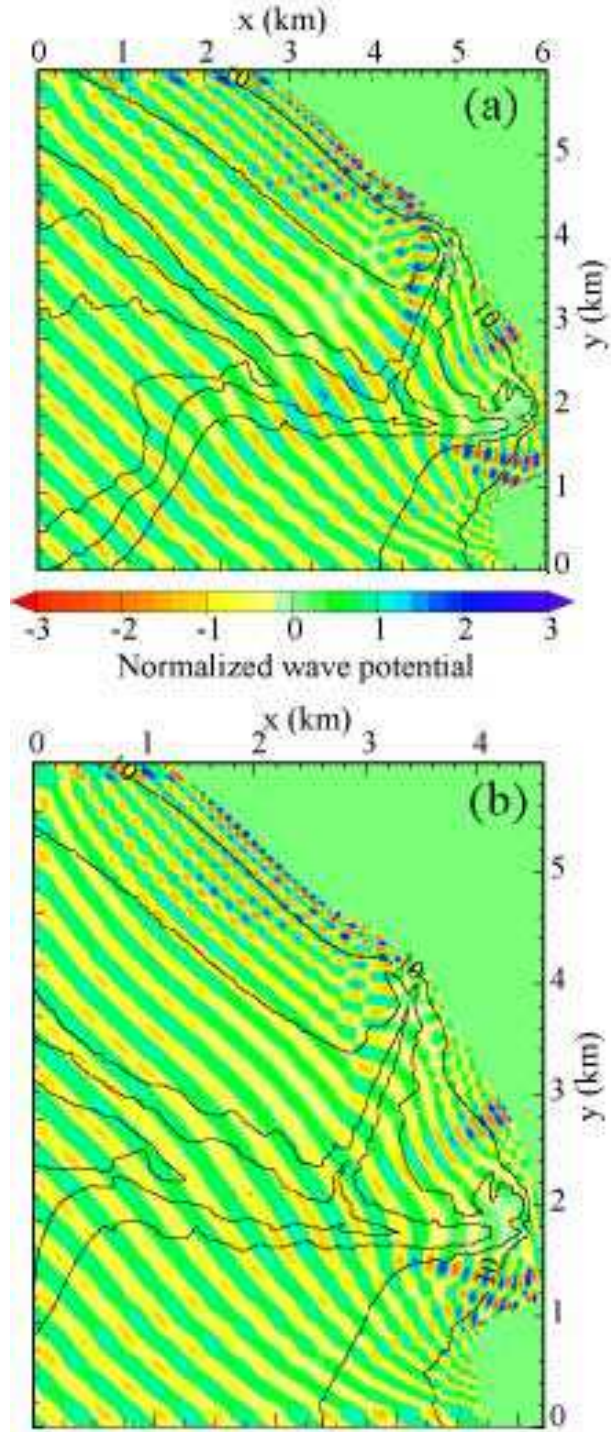}}
\caption{Computational domain for (a) $T>15$~s, and (b) $T\leq 15$~s. Also
shown are the NTUA5 solutions for the real part of the wave potential amplitude
for waves arriving from 270$^\circ$ with periods (a) $T=16$~s, and (b)
$T=15$~s, superimposed on the 10, 30, 100, 200, and 300~m depth contours.}
\label{Newdom2}
\end{figure}

The implementations of MSE, MMSE, NTUA5, and REF/DIF$1$ use two computational
domains with grids of 275 by 275 points (Figure \ref{Newdom2}). The larger
domain with a grid resolution of 21~m is used for wave periods longer than
15~s. The smaller domain, with a higher resolution of about 15~m, is used for
15~s and shorter waves. The $y$-axis of the grid is rotated 45$^{\circ}$
relative to North to place the offshore boundary in the deepest region of the
domain. Models were run for many sets of incident wave frequency and direction
($f$, $\theta$).
The CPU time required for one ($f,\theta$) wave component
calculation with the NTUA5 model (with $3$ evanescent modes) is about $120$~s
on a Linux computer with $2$Gb of memory and a 3~GHz processor. The wave
periods and offshore directions used in the computation range from $12$ to
$22$~s and $255$ to $340$ degrees respectively, with $0.2$~s and $2^\circ$
increments. The minimum period $12$~s corresponds to the shortest waves that
can be resolved with 7 points per wavelength in 10~m depth. Shorter waves are
not considered here because they may be affected by local wind generation, not
represented in the models used here, and are also generally less affected by
the bottom topography.

\begin{figure}[htbp]
\centerline{\noindent\includegraphics[width=20pc]{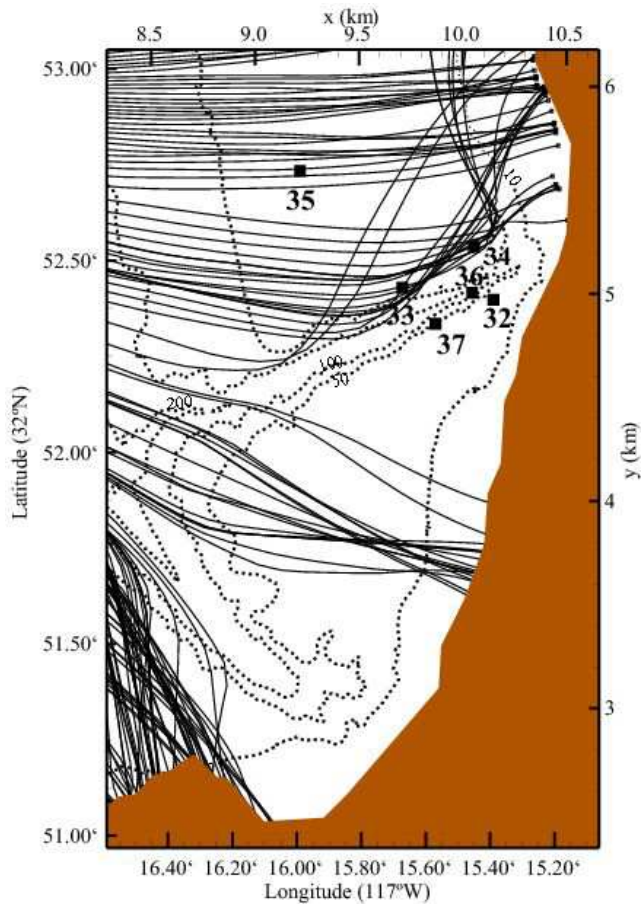}}
\caption{Location of directional wave buoys at the head of the Scripps canyon,
and wave rays for an offshore direction of 272$^\circ$ and a period of 15.4~s,
corresponding to a frequency just below the peak of the observed swell on
November 30. Contrary to the backward ray tracing model used for estimating the
wave spectrum at nearshore sites, rays were integrated forward from parallel
directions and equally spaced positions at 15~m intervals along the offshore
boundary at $x=0$, 10~km to the West of the buoys, practically in deep water. }
\label{wbuoysloc}
\end{figure}
Transfer functions between the local and offshore wave amplitudes were
evaluated at each of the buoy locations and used to transform the offshore
spectrum. The backward ray-tracing refraction model directly evaluates energy
spectral transfer functions between deep water, where the wave spectrum is
assumed to be uniform, and each of the buoys located close to the canyon, based
on the invariance of the wavenumber spectrum along a ray
[\textit{Longuet-Higgins}, 1957\nocite{Longuet-Higgins1957}]. A minimum of $50$
rays was used for each frequency-direction bin (bandwidth 0.005~Hz by 5
degrees), computed over the finest available bathymetry grid, with 4~m
resolution. The model is identical to the CREST model described by
\cite{Ardhuin&al.2001}, and validated by \cite{Ardhuin&al.2003b} on the U.S.
East coast. The energy source term set to zero here. This propagation-only
version of the model is also called CRESTp, and is similar to the model used by
\cite{OReilly&Guza1993} and \cite{Peak2004}. It was further validated on the
West coast of France [\textit{Ardhuin}, \nocite{Ardhuin2006a}2006].

\subsection{Model-Data Comparison}
Long swell from the west was observed on $30$ November $2003$, in the absence
of significant local winds. In the present analysis we use only data from
Datawell Directional Waverider buoys. The Torrey Pines Outer Buoy is
permanently deployed by the Coastal Data Information Program (CDIP), and
located about $15$ km offshore of Scripps Canyon. That buoy provided the deep
water observations necessary to drive the wave models. The directional
distribution of energy for each frequency was estimated from buoy measurements
of displacement cross-spectra using the Maximum Entropy Method [\textit{Lygre
and Krogstad}, 1986\nocite{Lygre&Krogstad1986}]. The NCEX observations were
made at six sites around the head of Scripps Canyon (figure \ref{wbuoysloc}).

All spectra used in the comparison, including the offshore boundary condition,
were averaged from 13:30 to 16:30 UTC, so that the almost continuous record
yields about 100 degrees of freedom for each frequency band with a width of
0.005~Hz. On that day the wind speed close to the coast did not exceed
3~m~s$^{-1}$, as measured by the CDIP Torrey Pines Glider port anemometer, and
the National Data Buoy Center (NDBC) buoy 46086, located 70 km West of San
Diego and representative of the entire modelled area.

\begin{figure}[htbp]
\centerline{\noindent\includegraphics[width=20pc]{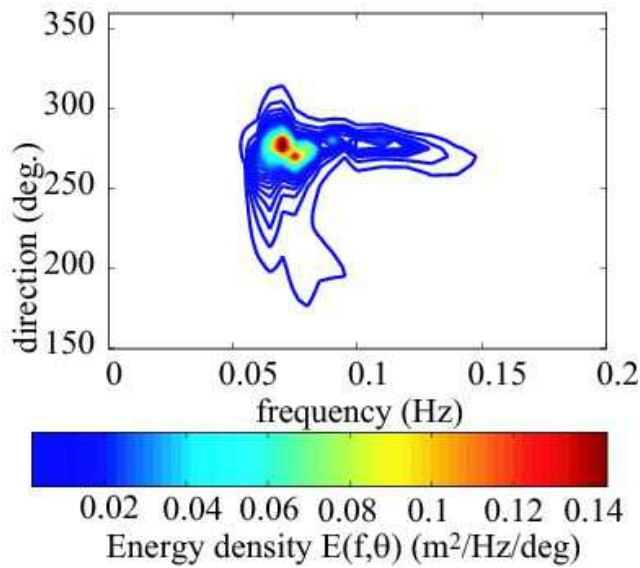}}
\caption{Directional wave spectrum  at Torrey Pines Outer Buoy at 15:00 UTC on
30 November 2003. } \label{Sp15h00}
\end{figure}

\begin{figure}[htbp]
\centerline{\noindent\includegraphics[width=20pc]{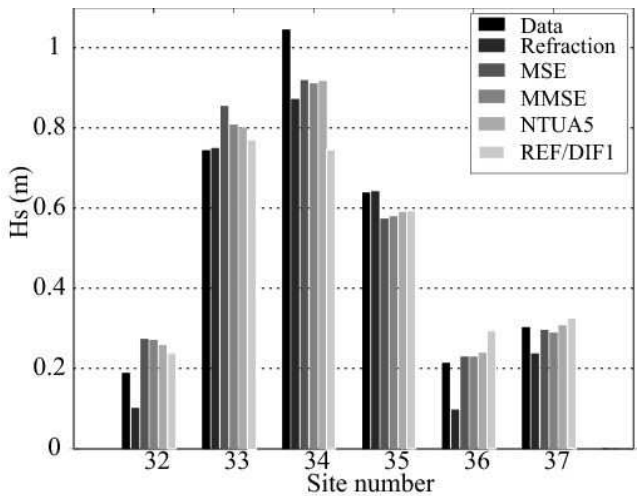}}
\caption{Comparison of predicted and observed significant wave height
($12s<T<22s$) for the 30 November 2003 swell event. Instrument locations are
shown in figure \ref{wbuoysloc}.} \label{wbuoysHs}
\end{figure}

The observed narrow offshore spectrum has a single peak with a period of $14.5$
s, and a mean direction of $272$ degrees, corresponding to an incidence angle
$\theta_i$ (relative to the Scripps Canyon axis) of $65^\circ$
(Figure\ref{Sp15h00}).
The model hindcasts are compared with observations  in Figure \ref{wbuoysHs}.
While the local amplification of the wave height at the head of canyon varies
with the incident wave direction, a dramatic reduction of the wave height
downwave of the rim of this canyon is predicted for all directions. Thus the
selected west swell case ($T_p=14.5s$, $\theta=272^{\circ}$) is representative
of the general wave transformation in this area, for low frequency swells
arriving a large range of directions. Significant wave heights $H_s$ were
computed from the measured and predicted wave spectra at each instrument
location, including only the commonly modelled frequency range ($f_1=0.05$~Hz,
$f_2=0.08$~Hz). The predicted $H_s$ is given by
\begin{equation}
H_s=4\left(\int_{f_1}^{f_2}\int_{\theta_1}^{\theta_2}M(f,\theta)
E(f,\theta)dfd\theta\right)^{1/2},
\end{equation}
where $E(f,\theta)$ is the observed offshore frequency-directional spectrum and
$M(f,\theta)$ is the model prediction of the ratio between the local and
offshore wave energies for the frequency $f$ and offshore direction $\theta$,
obtained by squaring the sea surface elevation transfer function.

\begin{figure}[htbp]
\centerline{\noindent\includegraphics[width=20pc]{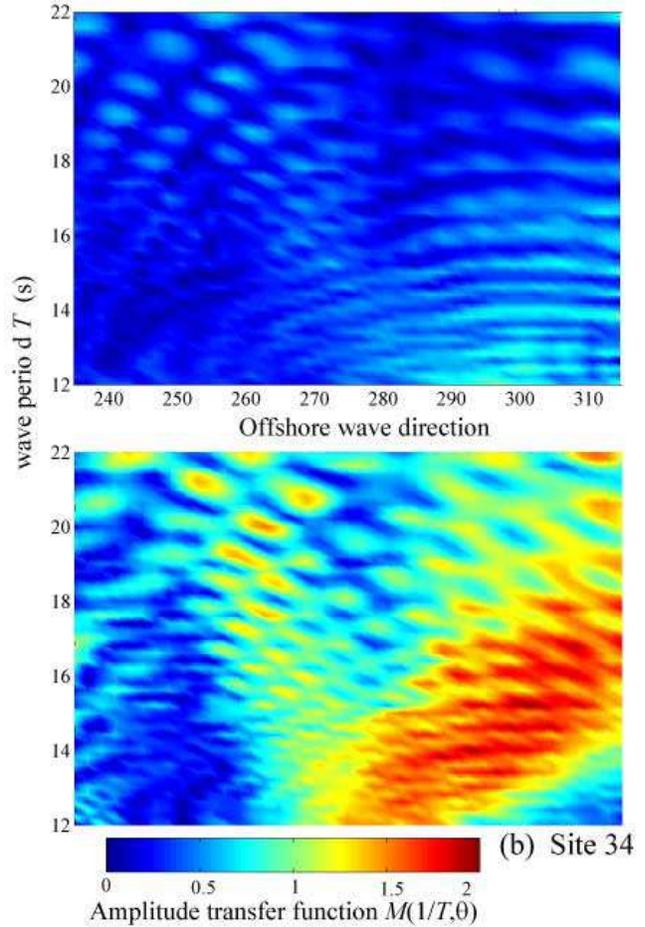}}
\caption{Amplitude transfer functions at site $34$ (a) and site $37$ (b),
defined as the ratio of the local and offshore wave amplitude modulus and
computed with NTUA $5$.} \label{TF3437}
\end{figure}

Observations show a dramatic variation in wave height across the canyon (figure
\ref{wbuoysHs}).
The offshore wave height is slightly enhanced at sites 33 and 34, in water
depths of 34 and 23~m respectively, along the north side of the canyon, and
slightly reduced on the shelf north of the canyon at site 35, in 34~m depth. A
dramatic reduction in wave heights is observed at sites 36, 37 and 32, over the
Canyon and on the south side, where the water depths are 111, 49 and 24~m,
respectively. Between buoys 34 and 36 the wave height drops by a factor 5 over
a distance of only 150 m, that is less than the 216~m wavelength at the peak
frequency (at the shallowest of the two sites). Such a pattern is generally
consistent with refraction theory as illustrated by forward ray-tracing in
figure \ref{wbuoysloc}. Whereas rays crossing the shelf north of the canyon
show the expected gradual bending towards the shore, rays that reach the canyon
northern wall are trapped on the shelf, and reach the shore in a focusing
region north of the canyon (Black's beach). From that offshore direction, and
an offshore ray spacing of 15~m, no rays are predicted to cross the canyon, so
that the south side of the canyon is effectively sheltered from 16~s Westerly
swells, in agreement with the observed extremely low wave heights (figure
\ref{wbuoysHs}, see also \cite{Peak2004}). The  amplitude transfer functions
$(M(f,\theta)^{1/2}$) are not overly sensitive to the wave frequency and
direction, as illustrated in figure \ref{TF3437}.a-b with NTUA5 predictions at
sites $34$ at the head of the canyon, and $37$ behind the canyon.

Up-wave of the canyon (instruments $33$, $34$, $35$), all models are found to
be in fairly good agreement with the observations. However, REF/DIF$1$
underestimates the wave height at site $34$. At this site, wave energy is
strongly focused by refraction, with rays turning by more that 90$^\circ$
(figure \ref{wbuoysloc}). The parabolic approximation does not allow such a
large variation in wave direction. Over and down-wave of the canyon
(instruments $32$, $36$, $37$), the wave heights predicted by MSE, MMSE and
NTUA5 agree reasonably well with the observations, whereas REF/DIF$1$ slightly
overestimates the wave height. For $f<0.06$~Hz few rays cross the canyon and
the energy predicted by the refraction model is extremely low, about $5$\% of
the offshore energy the total energy. This strong variation in wave energy
across the canyon is reduced by diffraction, which is not taken into account in
this refraction model, resulting in an under-prediction of the wave height at
the sheltered sites $32$, $36$, and $37$.

The sea state at that time also include an important contribution from higher
frequencies (figure 13). Significant wave heights computed over a wider
frequency range ($0.05<f<0.2$Hz), by adding the refraction model results to the
low-frequency results of other models,  vary little between the models, now
dominated by short wave energy.
\begin{figure}[htbp]
\centerline{\noindent\includegraphics[width=18pc]{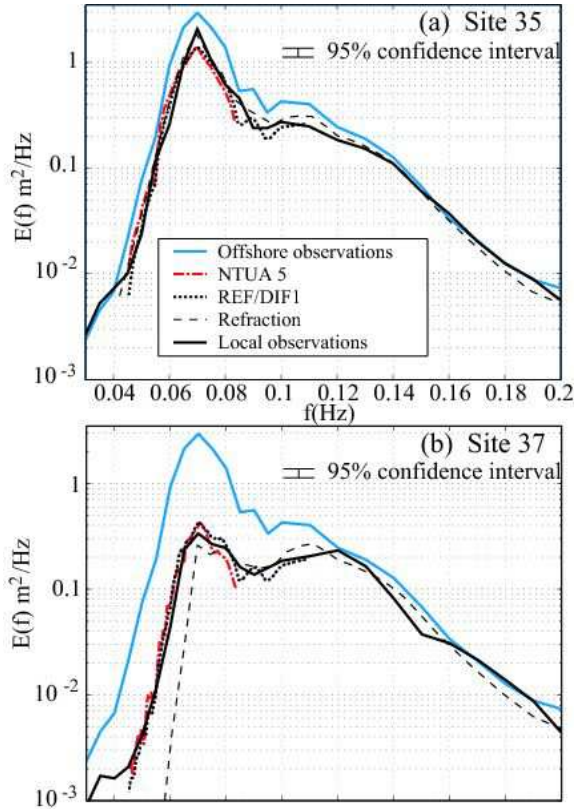}}
\caption{Comparison of predicted and observed frequency spectra at (a) site 35,
and (b) site 37, for the 30 November 2003 swell event.} \label{spectra35&37}
\end{figure}
\begin{figure}[htbp]
\centerline{\noindent\includegraphics[width=20pc]{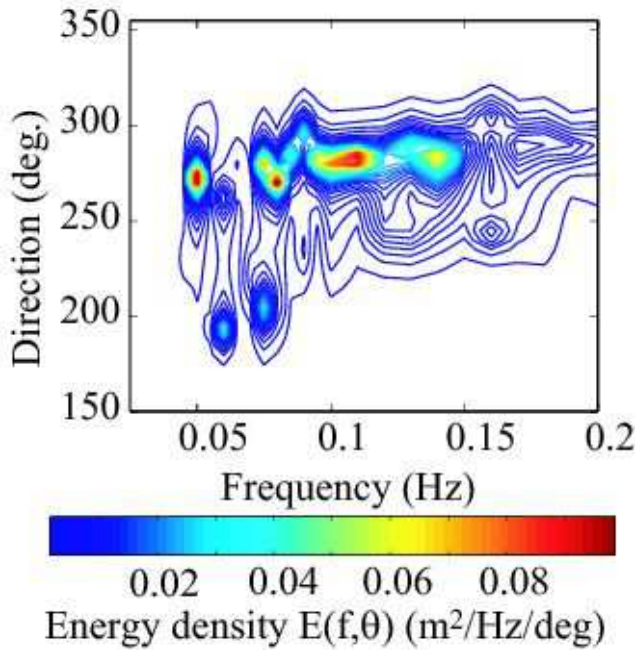}}
\caption{Directional wave spectrum  at Torrey Pines Outer Buoy at 12:00 UTC on
12 December 2003.} \label{Sp_28}
\end{figure}
However, wave heights are still markedly different between the buoys. It thus
appears that refraction plays an important role for frequencies up to 0.14~Hz
(see the difference in offshore and local spectra on figure 13), while
diffraction effects are significant, in that area, only up to 0.07~Hz. Further
confirmation of the trapping of low frequency waves is provided by another case
observed on $12$ December $2003$ (Figure \ref{Sp_28}), which we analyze with
the same method. The observed spectra are averaged from 12:00 UTC to 15:00
UTC.The observed spectrum has three peaks with a period of $20$, $12.5$ and
$9$~s, a mean direction of $270$, $270$ and $285$ degrees respectively and a
significant wave height of $1.9$m.
\begin{figure}[htbp]
\centerline{\noindent\includegraphics[width=20pc]{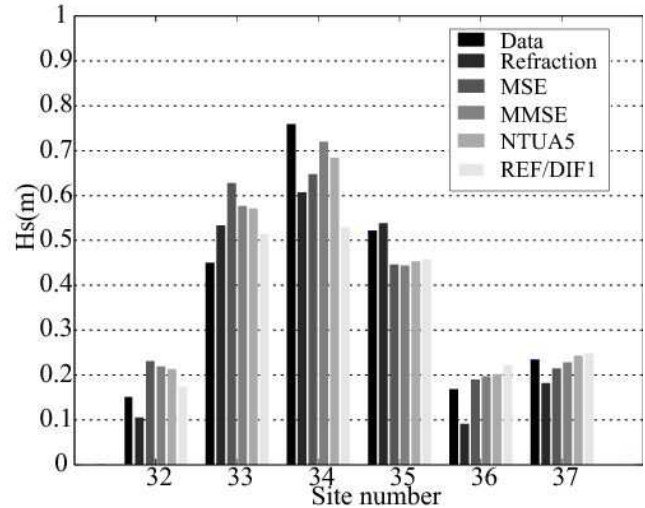}}
\caption{Comparison of predicted and observed significant wave height
($12s<T<22s$) for the 12 December 2003 swell event. Instrument locations are
shown in figure \ref{wbuoysloc}.} \label{wbuoysHs_12_12}
\end{figure}
The model hindcasts are compared with observations  in Figure
\ref{wbuoysHs_12_12}. Significant wave heights $H_s$ were computed from the
measured and predicted wave spectra at each instrument location, including only
the commonly modelled frequency range ($f_1=0.05$Hz, $f_2=0.08$Hz). On that day
the wind speed did not exceed 7~m~s$^{-1}$, as measured by the CDIP Torrey
Pines Glider port anemometer, but reached 13.5~m, blowing from the North West,
at NDBC buoy 46086. Such a wind is capable of generating a local wave field
with frequencies down to 0.095~Hz for fully-developed wave conditions.

\begin{figure}[htbp]
\centerline{\noindent\includegraphics[width=20pc]{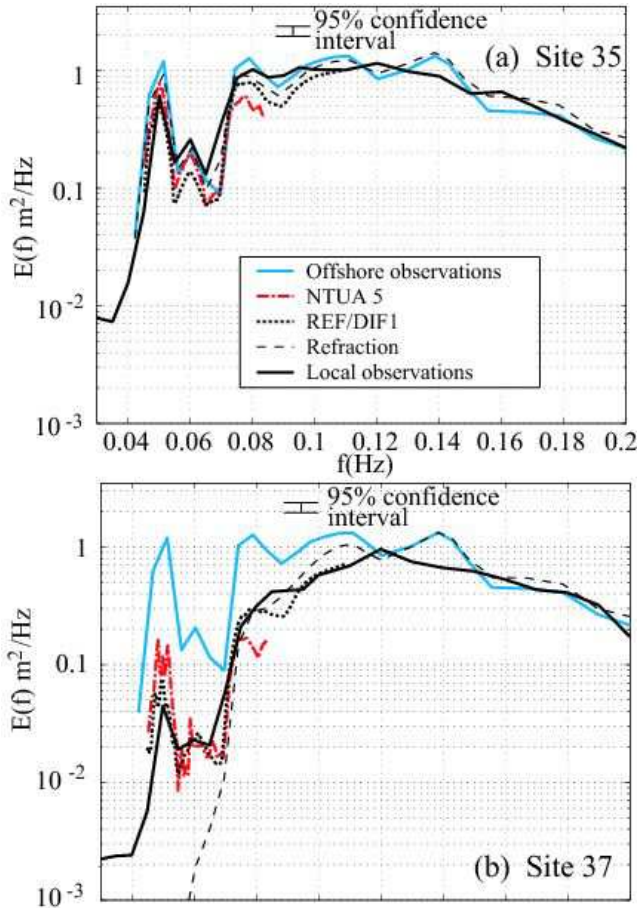}}
\caption{Comparison of predicted and observed frequency spectra at (a) site 35,
and (b) site 37, 12 December 2003 swell event.} \label{spectra35&37_28}
\end{figure}

As in the previous case, a large variation in wave height was observed across
the Canyon (Figure \ref{wbuoysHs_12_12}). Again, that variation remains limited
to a factor 10 difference for any wave frequency (compare Figure
\ref{spectra35&37_28}a and b), whereas the geometrical optics approximation
predicts much larger gradients. We note a general agreement of the predicted
wave height by the models, with an underestimation of the refraction model for
sites located down-wave of the Canyon. The predicted frequency spectra are
represented on Figure \ref{spectra35&37_28}a,b at sites $35$ and $37$. At site
$35$, located up-wave of the Canyon wall, NTUA5 and REF/DIF1 models are in a
good agreement with the measurement for the low frequency peak ($0.05$ Hz), but
underestimate the $0.08$Hz peak. The refraction model overestimates the low
frequency peak, but is in good agreement with the $0.08$Hz peak. At site $37$,
located down-wave of the Canyon, NTUA5 and REF/DIF1 predict a strongly
attenuated low frequency peak, as is observed, whereas the refraction model
predicts no energy transmission across the canyon. Below a cut-off frequency of
about $0.065$~Hz, the canyon acts as a complete barrier in the geometrical
optics approximation. The energy in the second peak at $0.08$~Hz is only
reduced by a factor 4 across the canyon, an effect well described by all
models, and thus attributable to refraction. All models generally agree with
the observations for $0.07<f<0.2$, within the spectrum measurement confidence
interval,  except for an overestimation of the refraction model for the high
frequency peaks ($0.11$ and $0.14$ Hz) of the spectrum. However, due to the
local wind sea generation between the offshore buoy and locations around the
canyon, these propagation models are not reliable for $f>0.095$~Hz.

In the two events most of the wave evolution is accounted for by refraction.
However,  diffraction is included in the models based on the MSE and its
extensions, and this effect allows for a tunnelling of wave energy across the
canyon. In these models, wave heights across the canyon are thus larger, in
better agreement with observed wave heights and wave spectra at the sheltered
sites 32, 36 and 37 (figures \ref{wbuoysHs}, \ref{spectra35&37},
\ref{wbuoysHs_12_12}).

The differences between NTUA5, MSE and MMSE model predictions are very small
and thus only NTUA5 results are shown in figure \ref{spectra35&37}. It may
appear surprising that the wave height behind the canyon is still 20\% of the
offshore wave height whereas the 2D simulations with comparable incidence
angles yield wave heights much less than 5\%. However, the Scripps Canyon is
neither infinitely long nor uniform along its axis. The three-dimensional
topography apparently reduces the blocking effect of long period swells that
was found over two-dimensional canyons.

\section{Summary}
Observations of the evolution of swell across a submarine canyon obtained in
the nearshore canyon experiment (NCEX), were compared with predictions of
refraction and combined refraction-diffraction models including the
coupled-mode model NTUA5 valid for arbitrary bottom slope
[\textit{Athanassoulis and Belibassakis}, 1999; \textit{Belibassakis et al.},
2001]. Predictions of a spectral refraction model are in good agreement with
observations [see also \textit{Peak}, 2004 for the entire
experiment]\nocite{Peak2004}, demonstrating that refraction is the dominant
process in swell transformation across Scripps Canyon. The geometrical optics
approximation, on which the refraction model is based, turned out to be very
robust. Accurate spectral predictions were obtained with taht model even in
cases where the wave energy changes by a factor of 10 over three quarters of a
wavelength.

For waves longer than 12~s, even larger gradients are predicted by the
refraction model, but these gradients are not observed. At those frequencies,
accurate results were obtained with the NTUA5 model and elliptic mild slope
equation models that include diffraction, which acts as a limiter on the wave
energy gradients. Differences between the models were clarified with 2D
simulations using representative transverse profiles of La Jolla and Scripps
Canyons, showing the behavior of the far wave field as a function of the
incidence angle. The underestimation by the refraction model may be interpreted
as the result of wave tunnelling, i.e. a transmission of waves to water depths
greater than allowed by Snel's law, for obliquely incident waves [see also,
\textit{Thomson et al., 2005}]. This tunnelling effect cannot be represented in
the geometrical optics approximation, and thus the refraction model predicts
that all wave energy is trapped for large incidence angles relative to the
depth contours, while a small fraction of the wave energy is in fact
transmitted across the canyon. Although different from the classical
diffraction effect behind a breakwater [e.g. {\textit {Mei}}
1989\nocite{Mei1989}], this tunnelling is a form of diffraction in the sense
that it prevents a sharp spatial variation of wave amplitude, and induces a
leakage of wave energy in areas forbidden by geometrical optics.

Observations were also compared with a parabolic
refraction-diffraction model that is known to be inaccurate for
large oblique wave directions relative to the numerical grid, and
is shown here to overestimate the amplitude of waves transmitted
across the canyon and underestimate the amplitude of waves focused
at the head of the canyon. Finally, depending on the bottom
profile and incidence angle, higher order bottom slope and
curvature terms (incorporated in modified mild slope equations and
NTUA5), as well as evanescent and sloping-bottom modes (included
in NTUA5) can be important for an accurate representation of wave
propagation over a canyon at small incidence angles. For large
incidence angles, that are more common for natural canyons across
the shelf break, the standard mild slope equation (MSE) gives an
accurate representation of the variations in surface elevation
spectra that is similar to that of the full NTUA$5$ model. Yet,
further analysis of NCEX bottom velocity and pressure measurements
may show that the MSE or other mild slope models may not
accurately represent near bottom wave properties, as also
discussed by \cite{Athanassoulis&al.2003}.

\begin{acknowledgments}
The authors acknowledge the Office of Naval Research (Coastal Geosciences
Program) and the National Science Foundation (Physical Oceanography Program)
for their financial support of the Nearshore Canyon Experiment.  Steve Elgar
provided bathymetry data, Julie Thomas and the staff of the Scripps Institution
of Oceanography deployed the wave buoys, and Paul Jessen, Scott Peak, and Mark
Orzech assisted with the data processing. Analysis results of the infragravity
wave reflections across La Jolla Canyon were kindly provided by Jim Thomson.
The authors also acknowledge anonymous referees for their useful comments and
suggestions.
\end{acknowledgments}

\bibliographystyle{agu}
\bibliography{wave}

\end{article}

\end{document}